\documentclass[openacc]{rsproca_new}

\usepackage{graphicx}
\usepackage{multicol}
\usepackage{float}

\usepackage{lineno}
\jname{rsif}
\Journal{J R Soc Interface\ }

\begin{document}

\title{Using neural ordinary differential equations to predict complex ecological dynamics from population density data}

\author{
Jorge Arroyo-Esquivel$^{1}$, Christopher A Klausmeier$^{1,2,3,4,5}$, Elena Litchman$^{1,2,3,4}$}

\address{$^{1}$Department of Global Ecology, Carnegie Institution for Science\\
$^{2}$W. K. Kellogg Biological Station, Michigan State University\\
$^{3}$Program in Ecology and Evolutionary Biology, Michigan State University\\
$^{4}$Department of Integrative Biology, Michigan State University\\
$^{5}$Department of Plant Biology, Michigan State University}

\subject{Ecosystem, Computational biology}


\corres{Jorge Arroyo-Esquivel\\
\email{jarroyoesquivel@carnegiescience.edu}}

\begin{abstract}

Simple models have been used to describe ecological processes for over a century. However, the complexity of ecological systems makes simple models subject to modeling bias due to simplifying assumptions or unaccounted factors, limiting their predictive power. Neural Ordinary Differential Equations (NODEs) have surged as a machine-learning algorithm that preserves the dynamic nature of the data \cite{chen_neural_2018}. Although preserving the dynamics in the data is an advantage, the question of how NODEs perform as a forecasting tool of ecological communities is unanswered. Here we explore this question using simulated time series of competing species in a time-varying environment. We find that NODEs provide more precise forecasts than ARIMA models. {We also find that untuned NODEs have a similar forecasting accuracy as untuned Long-Short Term Memory neural networks (LSTMs) and both are outperformed in accuracy and precision by EDM models. However, we also find NODEs generally outperform all other methods when evaluating with the interval score, which evaluates precision and accuracy in terms of prediction intervals rather than pointwise accuracy.}  We also discuss ways to improve the forecasting performance {of NODEs}. The power of a forecasting tool such as NODEs is that it can provide insights into population dynamics and should thus broaden the approaches to studying time series of ecological communities.

\end{abstract}



\maketitle


\section{Introduction}

Predicting the behavior or dynamics of an ecological system from our knowledge of its components is one of the main objectives of ecology \cite{houlahan_priority_2017}. One of the main areas of ecological prediction is near-term ecological forecasting, which has been developing rapidly thanks to increased data collection and computing power in the past decades \cite{dietze_ecological_2017,lewis_power_2023}. Near-term ecological forecasting is an important tool for ecological management, as predicting the impacts of management decisions allows making these decisions in a more informed manner \cite{dietze_iterative_2018,tulloch_ecological_2020}. Near-term ecological forecasting is also a tool that supports adaptive management by informing of different possible ecological outcomes \cite{galloway_supporting_2021}.

One of the many tools available for forecasting are dynamical models. Dynamical models provide an explicit structure to describe a time series, which can be used to unveil complex ecological dynamics that may not be forecast with purely statistical models \cite{luo_ecological_2011,laubmeier_ecological_2020}. In addition to their use in ecology, the use of dynamical models is a common practice in other fields such as physics \cite{perez-vieites_probabilistic_2018}, meteorology \cite{demaria_simplified_2009}, and epidemiology \cite{ionides_inference_2006}.

Dynamical models, however, {have} several important limitations. Ecological processes may occur on different spatial and temporal scales, and assessing what processes have to be incorporated into an ecological model to properly capture the time series dynamics is a difficult task \cite{perevaryukha_uncertainty_2011,getz_making_2018,mozelewski_forecasting_2021}. In addition, stochasticity plays a key role in ecological dynamics, which may lead to important deviations from a purely deterministic model \cite{bjornstad_noisy_2001,shoemaker_integrating_2020}. 

Finally, proper parameterization of dynamical models is generally a data-intensive process, especially for more complex models. Complexity in ecological data comes from different sources that can affect the performance of a forecasting tool. In addition to the inherent stochasticity of ecological processes, observing these processes is generally difficult, which leads to an increased observation noise \cite{mcclintock_unmodeled_2010,morrison_observer_2016,pennekamp_intrinsic_2019}. Ecological systems can have a high number of factors with an even higher number of interactions between them \cite{daugaard_forecasting_2022}. In addition to these sources of complexity, the amount of data required to effectively observe these complex dynamics in ecological datasets is not always available \cite{hsieh_extending_2008}. This has been observed to induce uncertainty and errors when training in other types of data \cite{hanczar_small-sample_2010,varoquaux_cross-validation_2018}.

Equation-free methods such as Empirical Dynamical Modelling (introduced in ecology by \cite{sugihara_nonlinear_1990}. For a more recent reference, see \cite{ye_equation-free_2015}) or Machine Learning methods such as Recurrent Neural Networks \cite{lapeyrolerie_limits_2023} have been proposed to overcome these limitations. However, the non-parametric nature of these methods creates a "black-box" model, where the underlying dynamics of the system being modelled are not tractable. This issue may limit the capabilities of these models to forecasting only. Although models that only perform forecasting may still provide important validations of ecological theory \cite{dietze_ecological_2017}, the intractability of the dynamics in the data may limit the insight that these tools can provide for the ecosystem being studied.

Neural Ordinary Differential Equations (NODEs, introduced in \cite{chen_neural_2018}) are a novel Machine Learning algorithm that learns the dynamics of a time series by using a neural network as the function of a differential equation. One of the main advantages of NODEs over other methods is that, by learning the differential equations that describe the time series, those equations can then be further analyzed using tools of dynamical systems theory to infer possible dynamical properties of the system being studied. This makes NODEs a type of ``grey-box'' model, where new insights can be gleaned from the data by combining the learning capabilities of a ``black box'' model such as neural networks and our {capability of analyzing ``white box'' model such as dynamical models using tools such as dynamical systems theory or system identification of nonlinear dynamics} \cite{sohlberg_grey_2008,negrini_system_2021}.

NODEs have already been used as a powerful tool in several applications such as forecasting power demand \cite{xie_neural_2019}, and modelling the dynamics of lithium batteries \cite{brucker_grey-box_2021} and single-cell transcriptomic dynamics \cite{chen_deepvelo_2022}. In ecology, NODEs have already been used to infer community interactions from time series data of two \cite{bonnaffe_neural_2021,frank_optimizing_2023} and three \cite{bonnaffe_fast_2022} species but have not been applied to study the dynamics of multispecies communities.

Hybrid approaches to forecasting are known to provide better performance by combining models with different strengths to overcome the limitations of each type \cite{hajirahimi_hybrid_2019}. In the case of NODEs, its learning of the dynamics can be further expanded using our knowledge of ecological theory through dynamical models. This is possible through a natural extension of NODEs called Universal Differential Equations (UDEs, introduced in \cite{rackauckas_diffeqfluxjl_2019}). UDEs combine the neural network being trained inside NODE with some known dynamics of the system. This extension of NODEs has been shown to perform better as a forecasting tool than the regular NODE framework in chaotic physics systems \cite{linot_stabilized_2023}.

While NODEs and UDEs have the potential of being powerful forecasting tools, and NODEs have already been used in ecology as a tool to study ecological dynamics, their forecasting power in ecological systems, especially in complex multispecies communities with nonequilibrium dynamics is still unexplored. In this study we explore how the NODEs and UDEs capabilities in forecasting species and community dynamics depend on different sources of ecological complexity such as community size and observation noise intensities, as well as different lengths of training data.
    
\section{Methods}
\subsection{Data Generation}
To generate the synthetic data used to evaluate the forecasting capabilities of NODE, we used the simple model presented in \cite{kremer_species_2017} to describe a community of species competing for a shared resource in a fluctuating environment with the default parameters used in it. {Although this model is simple, it is capable of producing complex dynamics such as multi-species limit cycles.} This model describes, for $i=1,\ldots,M$, the rate of change of each population density $n_i$ being governed by a per-capita growth rate $\mu_i$ depending on temperature explicitly modelled as a function of time, a trait $z_i$ which represents the temperature for optimal growth in the original model, and a resource $R(t)$, and a density and species independent death rate $m$. To prevent species densities to go to unreasonably small numbers, we introduced a small immigration rate parameter $\lambda$ set to $10^{-3}$. In equation form, this is written as

\begin{equation}\label{eq:synth_model}
    \frac{dn_i}{dt} = \left(\mu_i(t,z_i)R(t)-m\right)n_i+\lambda.
\end{equation}

{The growth rate $\mu_i$ is a Gaussian function with its mean being the optimal temperature trait $z_i$ and variance $\sigma_{\hbox{trait}}$. This function reaches its maximum $\mu_{\max}$ at the mean $z_i$. Temperature is a periodic function with period $\tau$, mean $T_{\hbox{mean}}$ and amplitude $T_{\hbox{amp}}$.}

\begin{equation}
    \mu_i(t,z_i) = \mu_{\max}\exp\left(\frac{(T(t)-z_i)^2}{\sigma_{\hbox{trait}}^2}\right)
\end{equation}
\begin{equation}
    T(t) = T_{\hbox{mean}}+T_{\hbox{amp}}\sin\left(\frac{2\pi t}{\tau}\right)
\end{equation}

{The available resource R(t) is the difference between the total resource in the system denoted by the parameter $R_{\hbox{tot}}$ and the resource being held in the bodies of the organisms in the populations.}

\begin{equation}
    R(t) = R_{\hbox{tot}}-\sum_{i=1}^Mn_i(t)
\end{equation}

We incorporate observation noise into our synthetic data by adding a random perturbation $\varepsilon_i$ sampled from a normal distribution with mean $0$ and standard deviation $\sigma n_i$ \cite{solow_fitting_1995}. Thus, our synthetic data is comprised of a time series $\hat{n_i}(t)$ given by

\begin{equation}
    \hat{n_i}(t) = \max(n_i(t) + \varepsilon_i(t),0).
\end{equation}

Our sample time series are the runs of this algorithm for a total of 100 time steps {evenly sampled at times $t_i = i$. A sample time series can be found in Supporting Information Figure S1}. To evaluate the capabilities of the NODE algorithms on different sizes and complexities of time series, we generate synthetic time series of different community sizes (number of species $M=10,40$) and different magnitudes of observation noise ($\sigma = 0,10^{-3},10^{-1}$). For each combination of community size and observation noise magnitude, we create {5} different time series with randomly sampled traits $z_i$ {uniformly sampled between $T_{\hbox{mean}}-T_{\hbox{amp}}$ and $T_{\hbox{mean}}+T_{\hbox{amp}}$}, and initial conditions {uniformly sampled and scaled such that their sum is smaller than $R_{\hbox{tot}}$}. These sampled traits and initial conditions are chosen in a similar manner as those in \cite{kremer_species_2017}.

\subsection{NODE algorithms} The Neural Ordinary Differential Equation (NODE) algorithm, introduced by \cite{chen_neural_2018} and used through this work consist of a neural network with inputs $X$ and parameters $\theta = (\theta_{ijk})$ denoted by $\mathrm{NN}(X;\theta)$ that describes a differential equation of the form

\begin{equation}\label{eq:NODE}
    \frac{dX}{dt} = \mathrm{NN}(X;\theta).
\end{equation}

{The function $\mathrm{NN}(X;\theta)$ consists on a series of hidden layers where, at the $i$-th hidden layer, its node $j$ has a value given by the nodes of the $i-1$-th layer (being the $0$-th layer the input layer). Each node $k$ is transformed using a nonlinear activation function $f_i$ and all transformed values are averaged using weights $\theta_{ijk}$.}. This process is repeated until the output layer is reached. The parameters of the neural network $\theta$ are trained in order for the training time series $\hat{X}$ to minimize a loss function $L(X,\hat{X})$ with the solution of Equation \ref{eq:NODE}.

A natural extension of the NODE algorithm is the Universal Differential Equation (UDE) algorithm, introduced by \cite{rackauckas_diffeqfluxjl_2019}. An UDE expands Equation \ref{eq:NODE} by incorporating some known dynamics of the system as a function $g(X)$ into the system. This leads to the UDE framework to describe the time series $X$ by the following differential equation

\begin{equation}\label{eq:UDE}
    \frac{dX}{dt} = \mathrm{NN}(X;\theta)+g(X).
\end{equation}

In the case of our synthetic time series, we choose the known dynamics to be the density, species independent death rate $m$ in Equation \ref{eq:synth_model}, i.e. $g(\hat{n_i}) = -m\hat{n_i}$. {In this case, we assume the value of $m$ is known by setting it to the value used for generating the synthetic time series ($m=1$).}

In addition to the different neural network architectures, we also explore how including or not including time as an input of our neural network affects the forecasting performance of our algorithms. The inclusion of time allows the modelling of the time series as a time-dependent dynamical system, which has been noted before as a key factor to unveil ecological dynamics \cite{hastings_timescales_2016,hastings_transient_2018}, and has been used in other forecasting tools, especially when the system being studied has some sort of environmental forcing (such is the case of our time series) \cite{cook_ecological_2010,miller_forecasting_2017}. {Notice that although we know the dynamics of these time series are externally forced by temperature, we consider the naive case where this external forcing is not properly understood by including time as an input and not the external force.}

For each architecture, we create two different types of models depending on their inclusion of time as an input. The time-agnostic algorithms will only receive population densities as inputs, while the time-aware algorithms will receive the time as an additional input.

Therefore, through this work, we present four different types of neural networks separated by their architecture type (NODEs or UDEs) and their model type (time-agnostic or time-aware).

\subsection{Training and evaluation of NODE algorithms} For each of our sample time series, we train neural networks of the different model architectures mentioned above, using mean-squared error (MSE) as our base loss function. To increase the generalization capabilities of the algorithms under observation noise, we add Lipschitz regularization into our loss function, as described in \cite{negrini_system_2021}.

To prevent rounding errors for population densities near 0, we train the neural networks using the logarithm of the time series. To test the capabilities of NODE in small data environments, we train our neural networks using different sizes of time series (10, 30, and 50 time steps).

Previous forecasting efforts with neural networks have found that testing performance of neural networks is highly sensitive to its initial parameters \cite{vidyarthi_modeling_2020,zhou_performance_2021,lapeyrolerie_limits_2023}. To properly test the forecasting capabilities of NODEs and UDEs, similar to \cite{lapeyrolerie_limits_2023}, we create 4 different neural networks of the same model architecture for each synthetic time series and test the performance of each model architecture based on the aggregate information of each different trained network. The development and training of these models is done using the DiffEqFlux.jl package in Julia 1.8 \cite{rackauckas_diffeqfluxjl_2019}.

For each of these neural networks, we test their performance by forecasting 50 time steps in the future from the trained dataset. We look at {three} standard performance metrics, the normalized RMSE of the prediction 10 time steps ahead, where the RMSE is normalized by the mean of the time series, the sizes of the prediction intervals, {and the interval score, which measures precision as the size of the prediction intervals and accuracy as how often the prediction interval contains the real observation \cite{bracher_evaluating_2021}}. To further test if NODEs and UDEs can properly infer ecological properties of the dynamics, we also test the power of the algorithms to predict the final community size of our time series. We measure this by calculating the number of species with high enough densities at the end of the forecasting window. We set this threshold of high enough densities to be 1, i.e. we count only species that satisfy $n_i>1$ at the end of the forecasting window. We then calculate the deviation from the final community size as the relative difference between the real number of species with population densities higher than 1 and the number predicted by our different models.

We compare these metrics with a null model, which in our case is the autorregressive integrated moving average (ARIMA) model. The ARIMA model is a standard, statistically sound model which is trained on the lags of the time series. ARIMA models are generally used as a point of comparison for other forecasting tools, and has been used before to evaluate the performance tools similar to NODEs and UDEs \cite{siami-namini_comparison_2018}. To create these ARIMA models, we use the auto.arima function of the forecast package in R 4.2 \cite{hyndman_automatic_2008}. We also compare our models with state-of-the-art, nonparametric algorithms in ecological forecasting, including Empirical Dynamical Modelling (EDM) \cite{munch_recent_2023} and Long Short-Term Memory (LSTM) \cite{lapeyrolerie_limits_2023}. We create EDM models in R using the rEDM package and LSTM models using TensorFlow using the default parameters for both algorithms.

The comparison between algorithms is statistically supported using a multiple-factor ANOVA, using the architecture and model types {(where ARIMA, LSTM, and EDM architectures are considered time agnostic)}, as well as the training dataset size, community size $M$ and observation noise $\sigma$ as factors. {To account for the dependence between samples coming from the same time series, we include time series identity as a random effect.} We do not determine a threshold of statistical significance, as this is not appropriate when working with simulated datasets \cite{white_ecologists_2014}. Instead, and similar to \cite{corell_depth_2012}, we study the importance of each factor in determining the value of each of the evaluated metrics by measuring their contribution to variation. This contribution to variation is calculated as the ratio of the sum of squares of each separate factor and the total sum of squares.

\section{Results}
\begin{figure}[H]
\centering
\includegraphics[width=\linewidth]{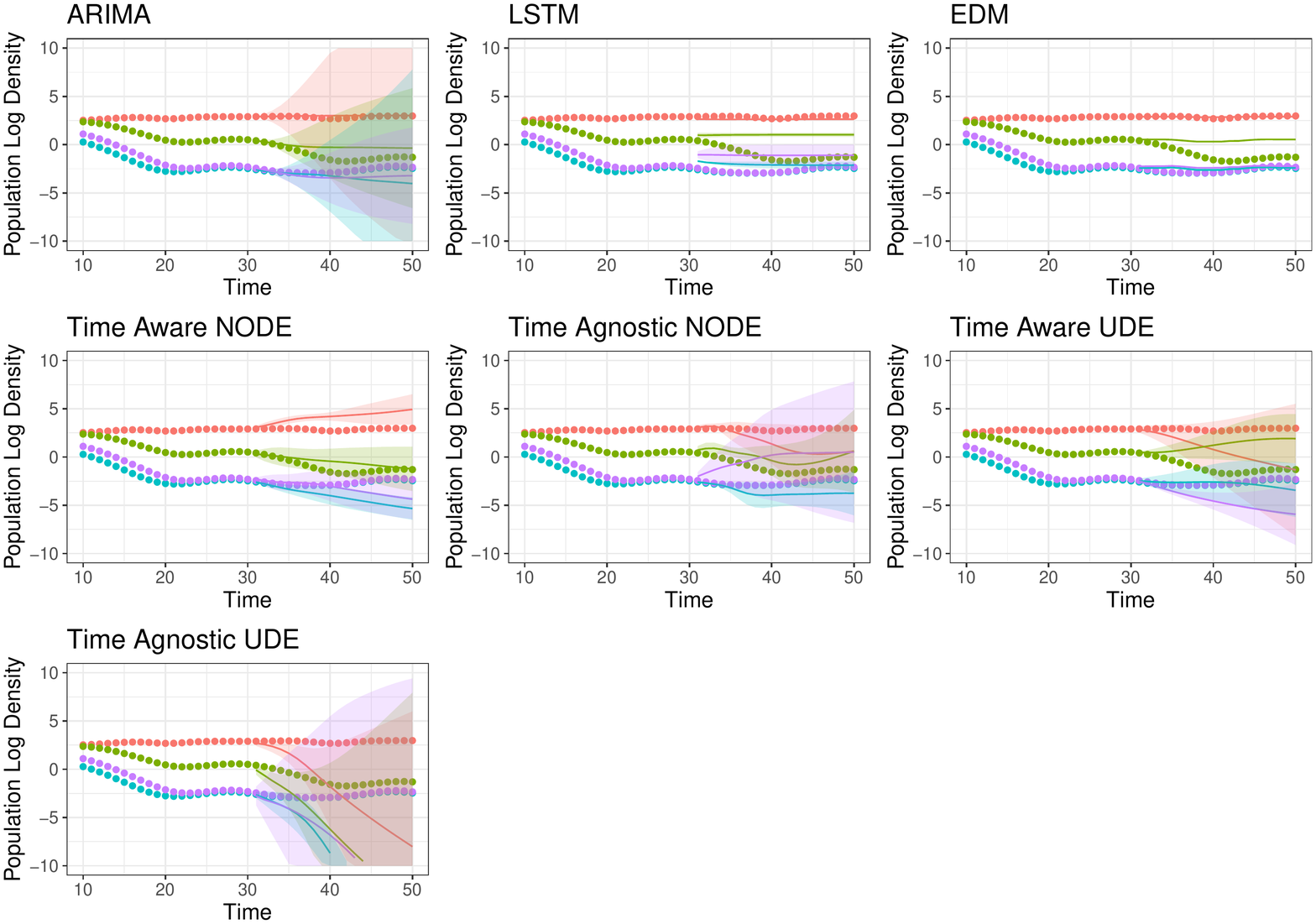}
\caption{Forecasting of a sample time series in log base 10 after training for 30 time steps with a forecasting horizon of 20 time steps with the different types of forecasting tools explored in this paper. For simplicity of visualization, we only show here 4 of a total of 10 species time series of population densities. In this figure, dots represent the synthetic data, and solid lines represent the forecasted time series. See Methods section for further details.}\label{fig:forecasts}
\end{figure}

From a sample time series forecasted with the different models we tested, several interesting properties arise (Figure \ref{fig:forecasts}). We observe that {NODE and UDE models follow the oscillatory behavior of the populations} but tend to predict an eventual collapse of the lower density populations. {ARIMA and EDM models follow the oscillatory dynamics of most of the populations, while LSTM models follow the dynamics more weakly and resemble a constant function in the logarithmic scale of the figure}. Visually, NODE and UDE models appear to have a smaller forecasting uncertainty than our null model, the ARIMA model. The prediction intervals of ARIMA models increase exponentially, whereas those of NODE and UDE models increase at a smaller rate, and, in some populations, follow the population dynamics before starting to increase. There also does not appear to be any noticeable difference in performance between time aware and time agnostic models. EDM and LSTM models have the smallest prediction intervals of all, although some of the shown populations tend to be overestimated, especially in the LSTM case.

\begin{figure}[H]
\centering
\includegraphics[width=\linewidth]{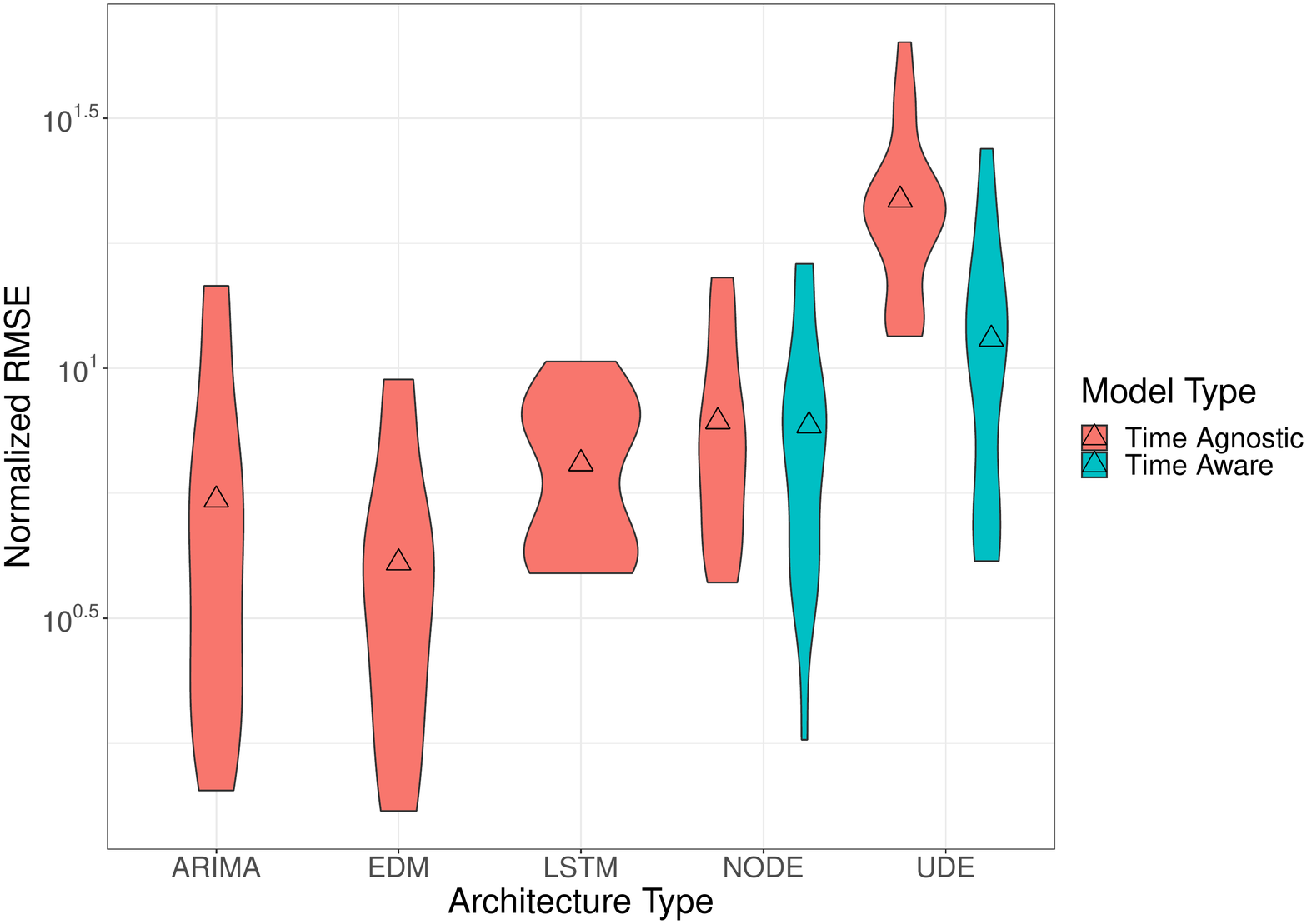}
\includegraphics[width=\linewidth]{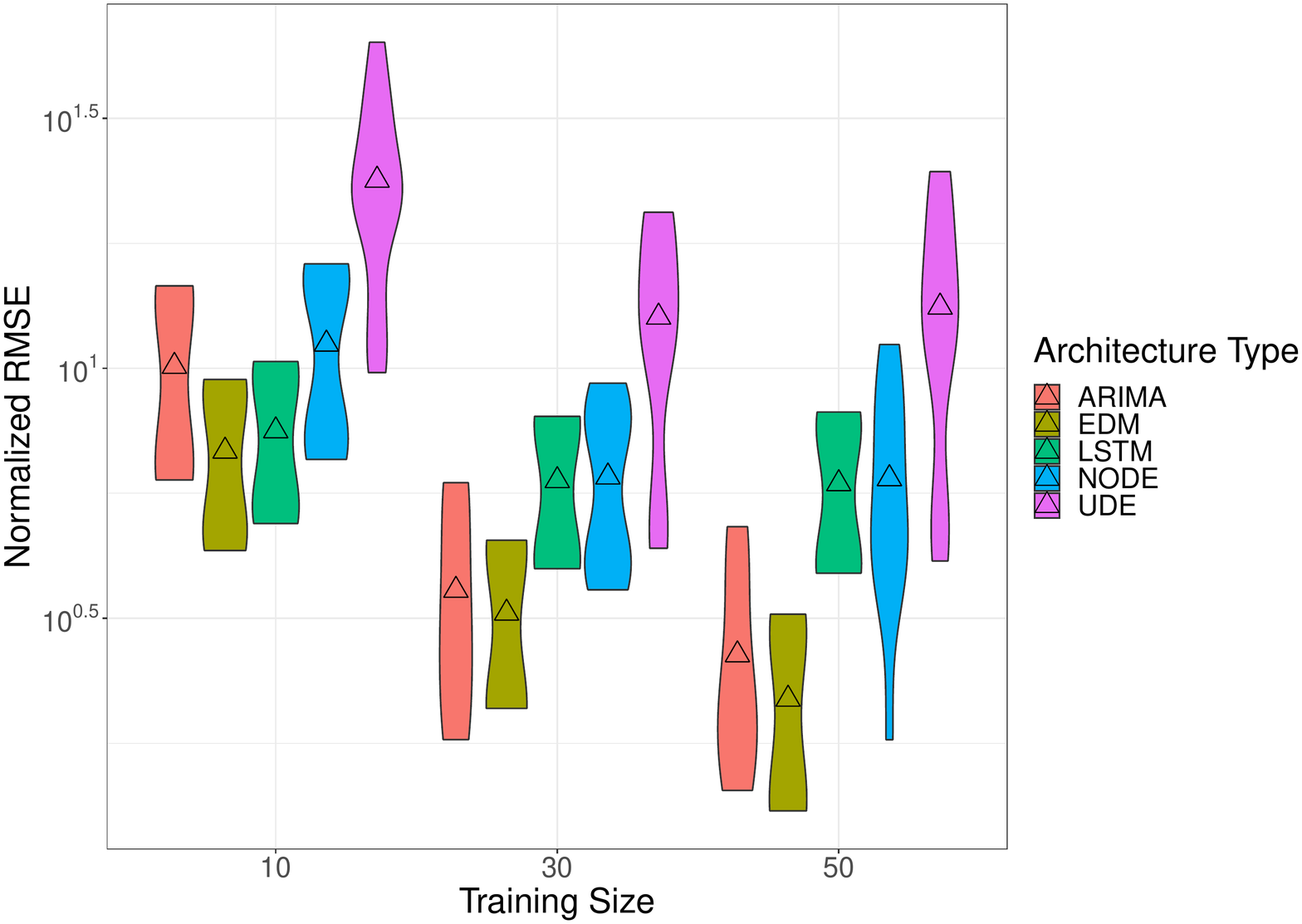}
\caption{Violin plot of normalized RMSE for future forecasts of the time series separated by the type of model architecture (top) and its training size (bottom) after a forecast horizon of 10 time steps. The triangles indicates the mean of the distribution.}
\label{fig:RMSE_violins}
\end{figure}

Measuring the forecasting performance of a wider set of models through the RMSE of the forecast shows that, visually, the mean RMSE of NODE models is close to that of ARIMA and LSTM models, with EDM being the more accurate forecasting tool (Figure \ref{fig:RMSE_violins}). UDE models, however, have a higher mean RMSE than NODEs and ARIMA, potentially suggesting that UDEs are worse forecasting tools than ARIMA, especially in the case where time is not included as an input.  We also find that all models benefit from an increased training size. However, LSTMs, NODEs, and UDEs did not having a noticeable increase in accuracy as the training size increased from 30 to 50 time steps of training. An ANOVA analysis supports these results, as the architecture type are among the variables with the biggest contribution to the variation observed in our data, and training size {contributes more than 60\% of the variation in the data} (Supporting Information Figure S2).

\begin{figure}[H]
\centering
\includegraphics[width=\linewidth]{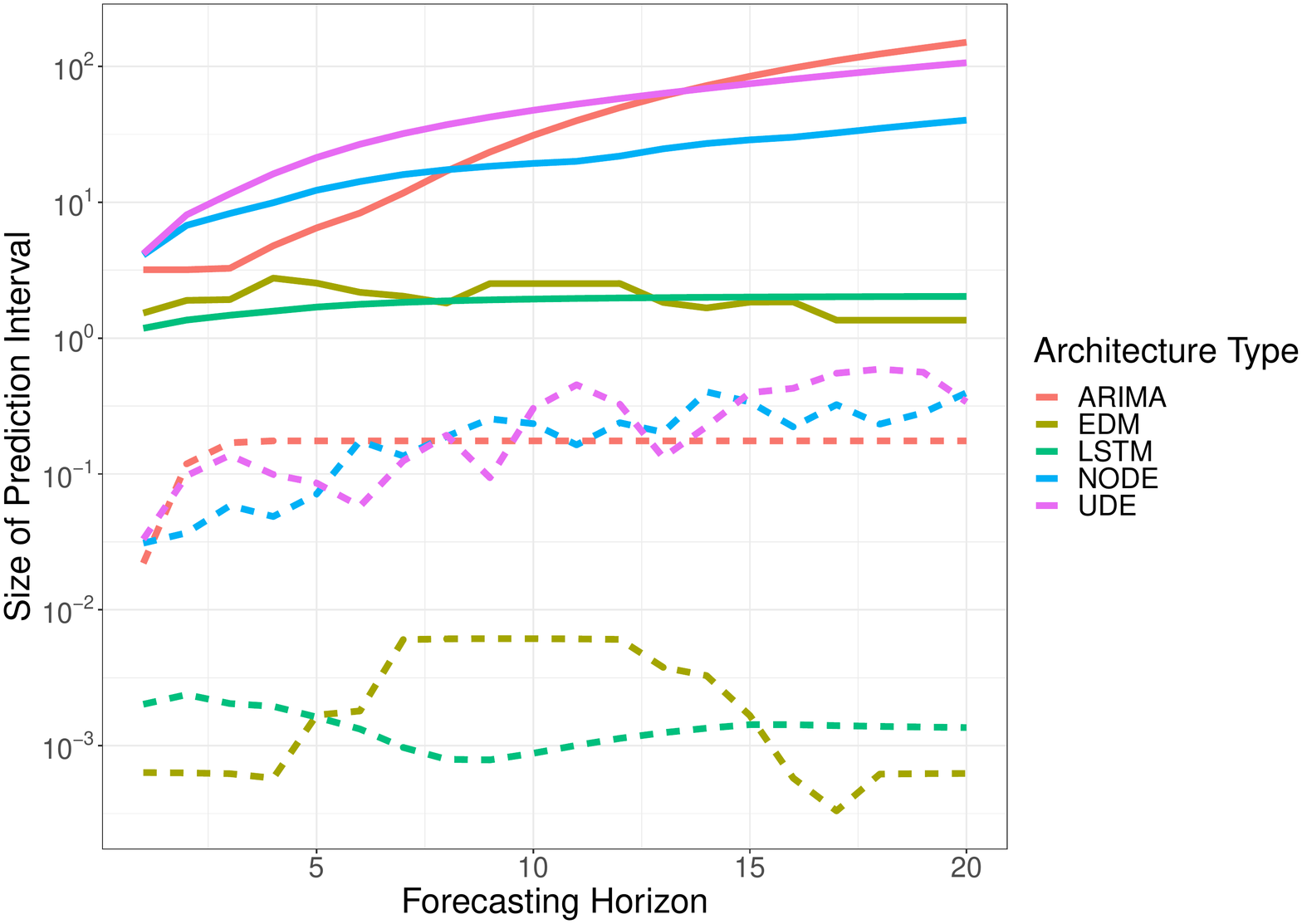}
\caption{Time series of the spread of uncertainty for future forecasts of the time series as the forecasting horizon increases, including the minimum (dashed lines) and maximum (solid lines) values of spread for different forecasting horizons. Data are separated by the type of model architecture. {Notice the spread of uncertainty does not grow monotonically as methods that follow the dynamics more closely may work better at some later forecasting horizons.}}
\label{fig:spread_violins}
\end{figure}

Although NODEs and UDEs do not necessarily improve on the forecasting capabilities of ARIMA assessed with the RMSE, their prediction intervals tend to remain smaller than those of ARIMA models for a longer forecasting horizon (Figure \ref{fig:spread_violins}). Our ANOVA analysis suggests that architecture type is the main contributor to the observed variation in spread (Appendix Figure S3), while the inclusion of type having a smaller contribution, which is also observed in Figure \ref{fig:spread_violins}. This supports the idea that NODEs and UDEs reduce the uncertainty of the predictions, especially at longer forecasting horizons.

\begin{figure}[H]
\centering
\includegraphics[width=\linewidth]{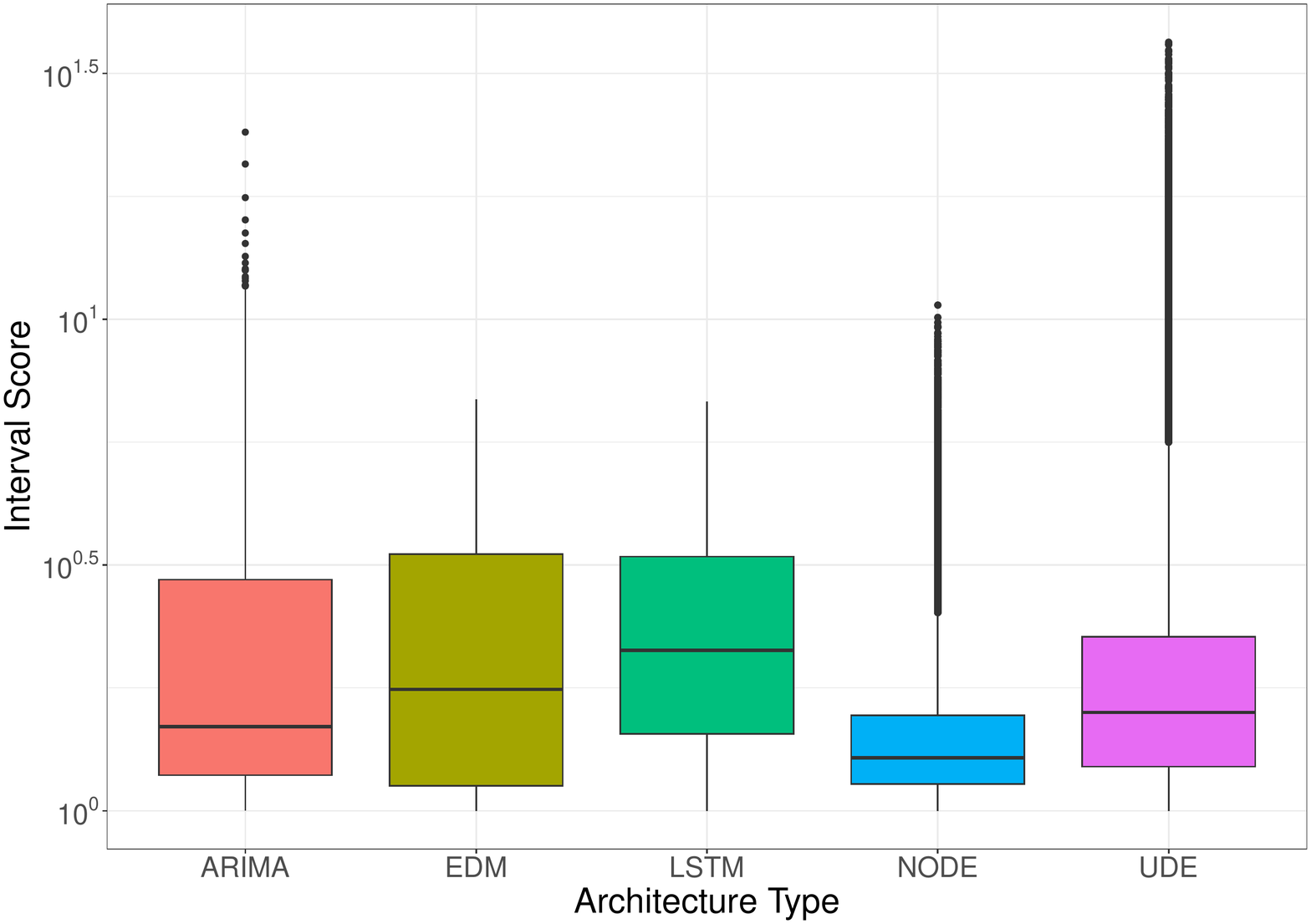}
\caption{Boxplot of the interval scores for future forecasts of the time series separated by the type of model architecture after a forecast horizon of 20 time steps.}
\label{fig:interval_scores}
\end{figure}

{When evaluating the performance of each architecture using the interval score, we find that NODEs generally outperform all other architectures (Figure \ref{fig:interval_scores}). This suggests that, although NODEs do not provide the most accurate mean prediction, their ability to follow the dynamics of the time series allows the forecast interval to contain the true value for longer while being precise enough. This is also supported by the ANOVA, where the architecture contributes the most to the variation in the data (Supporting Information Figure S3).}

\begin{figure}[H]
\centering
\includegraphics[width=\linewidth]{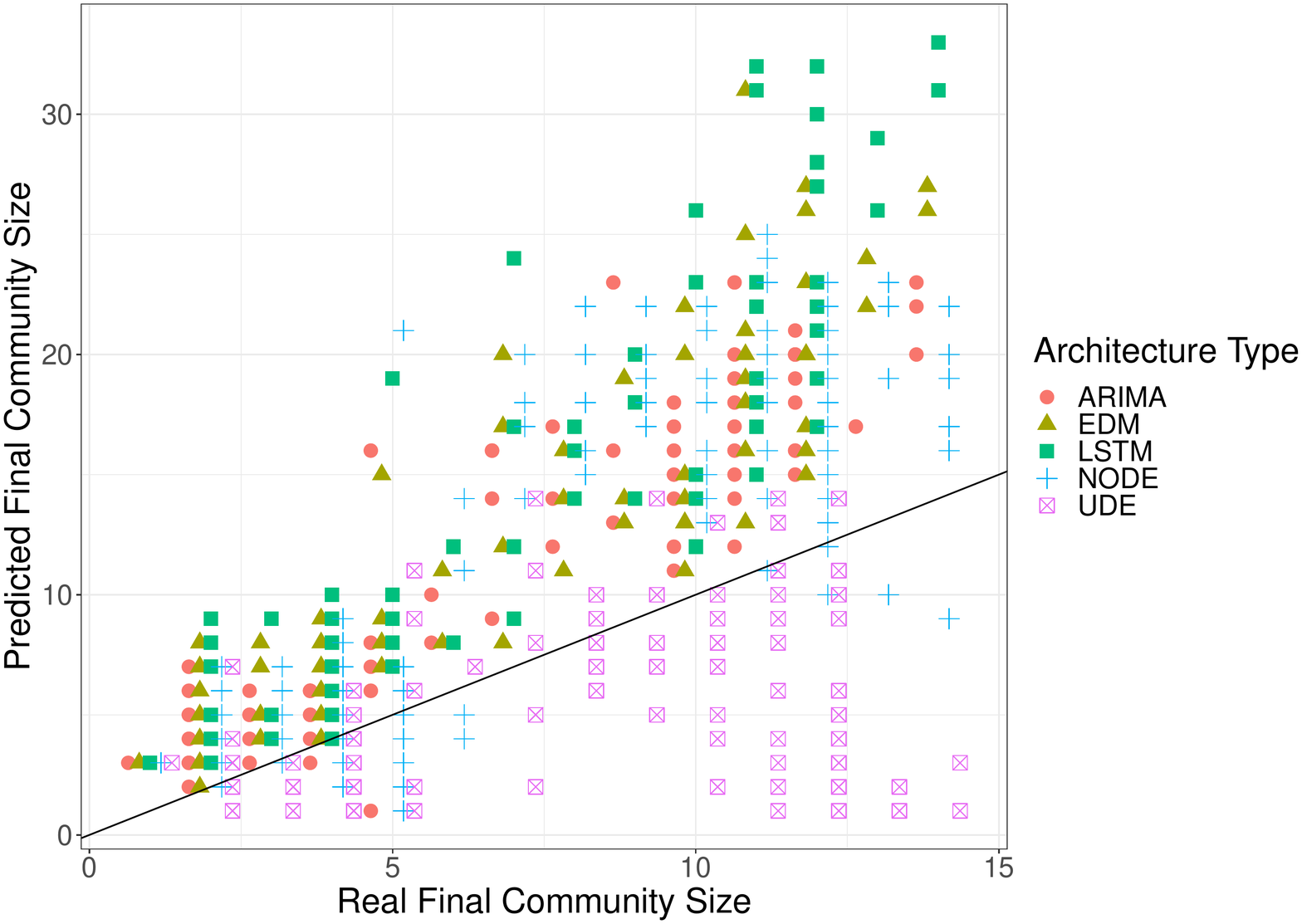}
\caption{Plot of real vs predicted of the final community size (number of species) after a forecast horizon of 50 time steps separated by the type of model architecture. The black line represents $x=y$ line, i.e., the forecasting tool predicted the same final community size (number of species) after the forecast horizon of 50 time steps.}
\label{fig:FCS_violins}
\end{figure}

By comparing the observed and predicted final community sizes, we find UDEs' predicted final community sizes tend to be closer to the real size than ARIMA and NODE models in the case of a smaller community size ($M = 10$, Figure \ref{fig:FCS_violins}). In the case of a bigger community size ($M = 40$), UDEs tend to underestimate the final community size, whereas the rest of the models tend to overestimate it.This is also supported by our ANOVA analysis, which shows that the architecture type is the variable that contributes the most to the variation in our data {besides community size, which had an important effect as mentioned before (Supporting Information Figure S4).}

\section{Discussion}

In this manuscript we {introduce} the forecasting capabilities of NODEs and UDEs{ in an ecological context} using {simulated} ecological time series data of multispecies communities. We found that overall prediction capabilities of simple NODEs and UDEs are comparable to that of ARIMA models. However, these models present two important improvements over ARIMA models. First, learning of the dynamical properties of the time series allows these models to reduce the uncertainty of the forecasts. Under equal forecasting accuracy, a less uncertain forecast is generally better, and the benefits of less uncertain forecasts of ecological dynamics have been known for several decades \cite{araujo_reducing_2005}. 

Second, UDEs showed an improvement over ARIMA at recovering the number of species at detectable densities at equilibrium at smaller community sizes. These results are consistent with those of \cite{linot_stabilized_2023}, where they found that UDEs with an explicit linear term were able to recover several statistical properties of a chaotic attractor in physical models. In higher community sizes, UDEs tended more to an underestimation of the final community sizes. This is potentially caused by our choice of known dynamics function $g(\hat{n_i}) = -m\hat{n_i}$, which induces a bias towards a decrease in population densities that may have been taken by the regularized neural network during training as a sign that more populations should have smaller densities. This could also explain the reduced forecasting accuracy of UDEs compared to other models, as well as why incorporating time as an input improved the accuracy of UDEs. This could be addressed in a future work using alternative training routines such as the multiple shooting method {which trains the NODE model starting from different initial conditions found in the data} \cite{turan_multiple_2022}. Other studies have suggested that regularization of the neural network is not necessary to prevent overfitting, as the known dynamics of the UDE implicitly regularize the neural network \cite{bolibar_universal_2023}. Whether this statement can be generalized or depends on the choice of known dynamics is a question to explore in future work.

The result of being able to predict the final community size is more impressive considering that training size was a small contributor of the variance observed in the simulated data (Supporting Information Figure S4). This includes the possibility of training only with transient data (the case with a training size of 10 time steps), showing that properties of the attractor can be recovered even without having observed it. This has several potential useful applications for ecological time series, such as identifying vulnerable populations from time series by inferring their extinction risk \cite{purvis_predicting_2000}, the presence of long transients in data \cite{reimer_noise_2021}, or inferring a possible regime shift \cite{scheffer_generic_2015}. This recovery of properties of the attractor also has the potential to make UDEs an informative tool of long-term forecasting, which has important benefits on decadal-scale ecological management \cite{ojaveer_ecology_2010,weng_relative_2011}.

Heuristically, the idea of UDEs being better at recovering dynamical properties of the time series goes back to the nature of hybrid models. Hybrid models allow for different modelling methods to complement each other and overcome the limitations of each model type \cite{hajirahimi_hybrid_2019}. In our case, the explicit dynamics provided to the UDEs facilitate the training of the neural networks and give them a bias to model certain dynamics that are expected by ecological theory. Other hybrid models in ecology have focused on the integration of agent-based models to unveil emerging properties from individual interactions \cite{parrott_hybrid_2011}. The implicit learning process of neural networks allows them to overcome several limitations of agent-based models, such as rule-setting or the computational difficulties in scalability of these models \cite{dsouza_framework_2008}. {This in turn shows that UDEs have the potential to be the partially specified ecological models of today \cite{wood_partially_2001}. Partially specified ecological models are semiparametric ecological models where a component of the dynamics are defined by a nonparametric estimator (such as a neural network). When the concept of partially specified ecological models was proposed, computational capacity limited what estimators could be realistically used. Modern technology allows a properly designed neural network and training scheme to be realistically used to learn unknown dynamics and create a better forecasting tool.}

Unlike UDE models, where the time inclusion improved forecasting accuracy, we found no effect of including time as an input in our NODE. This is an unexpected result, especially when considering that the model used to generate our time series includes a time-dependent environmental forcing process. One of the possible causes is that time is incorporated both as an input and as an output. This makes time susceptible to propagation of errors in the forecasting process. In the case of using this neural network architecture as a forecasting tool, time does not provide additional forecasting power, and instead the added complexity of an additional input and output unnecessarily increases computational time. Moreover, inclusion of time as an input and output limits one of the main strengths of NODE architectures over other time series analysis neural networks, which is the ability to learn the dynamics using unevenly sampled data. Encoding time as an input but not as an output requires a different neural network architecture than the one used in this paper (see e.g. \cite{davis_time_2020}), and will be explored in future work.

When comparing NODEs to other state-of-the-art algorithms, we find that NODEs perform similarly in accuracy and are outclassed in precision to LSTMs, and are outclassed in accuracy and precision by EDMs. {Although this is the case, we also find NODEs generally outperform both LSTM and EDM models .} LSTMs are known to improve their forecasting performance when compared to ARIMA models \cite{siami-namini_comparison_2018}. The fact that LSTMs with default settings performed similarly to NODEs suggests that a fine-tuned NODE model could outperform ARIMA models as well. How to fine-tune a NODE model to improve forecasting accuracy is an open question. EDMs have been used in ecology for several decades due to their forecasting power \cite{munch_recent_2023}. 

Although our results suggest that EDMs are a {more precise and accurate} forecasting tool, the compromise of having a ''grey-box'' model such as {NODE that can be used to discover dynamics approach may be useful in some situations. NODEs have been previously used in ecology with the purpose of gaining insight of the ecological dynamics of a time series \cite{bonnaffe_neural_2021,bonnaffe_fast_2022}.} More realistically, a forecasting tool will consist of an ensemble of different models, where the forecasting strenghts of EDM and the interpretability of NODE can be combined to produce better forecasts \cite{clark_near-term_2022}.

Our analysis focused on using neural networks to learn deterministic dynamics. Ecological time series are susceptible to environmental and demographic stochasticity, which may make standard NODEs prone to decreased forecasting performance due to process error \cite{lande_stochastic_2003}. In such a case, Neural Stochastic Differential Equations provide an extension that can improve by learning the stochastic noise in addition to the deterministic dynamics \cite{innes_differentiable_2019}.

Although we found that the forecasting accuracy in NODEs and UDEs is not necessarily better than the ARIMA model, there are several optimization techniques that can be used to improve the forecasting performance. For example, augmenting the input of the NODEs algorithms by adding empty dimensions \cite{dupont_augmented_2019} or explicitly modelling observation noise \cite{oganesyan_stochasticity_2020} has been shown to improve predictive capabilities of NODEs. Observation noise and other sources of uncertainty can be implicitly modelled through the Bayesian estimation of the weights of the neural network through an extension of NODEs called Bayesian NODEs \cite{dandekar_bayesian_2021}. In the case of UDEs, it is an open question whether providing a more detailed set or more complex known dynamics increases the forecasting performance of UDE models.

In addition to technical improvements of the neural network design, several domain-specific improvements can be done as well. In our model architectures, we only included population densities and time as inputs. However, ecological forecasting can incorporate other sources of data such as environmental information \cite{williams_improving_2009,richardson_estimating_2010,keenan_rate_2013} or population traits \cite{litchman_trait-based_2008,buckley_functional_2012,green_trait-based_2022} to improve forecasting performance. These covariates provide further information on the underlying ecological processes, which leads to a more informed statistical model \cite{brodie_trade-offs_2020}. How to properly incorporate these sources of data into a NODE framework is still an open question. However, using a hidden layer to encode this information into the population densities is a possible way which has been used before to aggregate different sources of data into a NODE framework \cite{wang_predicting_2022}.

{In this work we introduced the algorithms of NODEs and UDEs as forecasting tools in an ecological context. Our results suggest that, similar to other novel forecasting tools, fine-tuned NODEs and NODE-derived algorithms can provide unique benefits (such as an increased interpretability) when used individually or as part of a suite of ensemble models \cite{oidtman_trade-offs_2021,chowell_ensemble_2022}. An important limitation of our analysis is that training NODEs and UDEs is computationally expensive. This makes increasing the nummber of time series being trained, as well as iteratively training a single time series a slow process. As noted above, several questions regarding the optimization of NODE algorithms remain, including faster training algorithms, and a deeper exploration of its performance through a bigger synthetic dataset and with a real dataset. Addressing the many unique challenges ecological datasets have when applying NODEs and related methods with these and other explorations will lead to interesting theoretical, methodological, and ecological questions.}

\section*{Acknowledgements}
{Arroyo-Esquivel acknowledges the support of NSF Award No. 2233982: ``Model Enabled Machine Learning (MnML) for Predicting Ecosystem Regime Shifts''. Klausmeier and Litchman acknowledge the support of NSF Awards No. 1754250 ``Intraspecific trait variation in phytoplankton at different scales'' and No. 2124800: ``MIM: Using multilayer interaction networks to predict microbiome assembly and function''. We would also like to acknowledge Drs Jonas Wickman and Veronica Frans for providing valuable comments on early drafts of this paper.}

\printbibliography
\appendix

\renewcommand{\figurename}{Figure S}
\setcounter{figure}{0}
\section*{Supporting Information}
\begin{figure}[H]
\centering
\includegraphics[width=\textwidth]{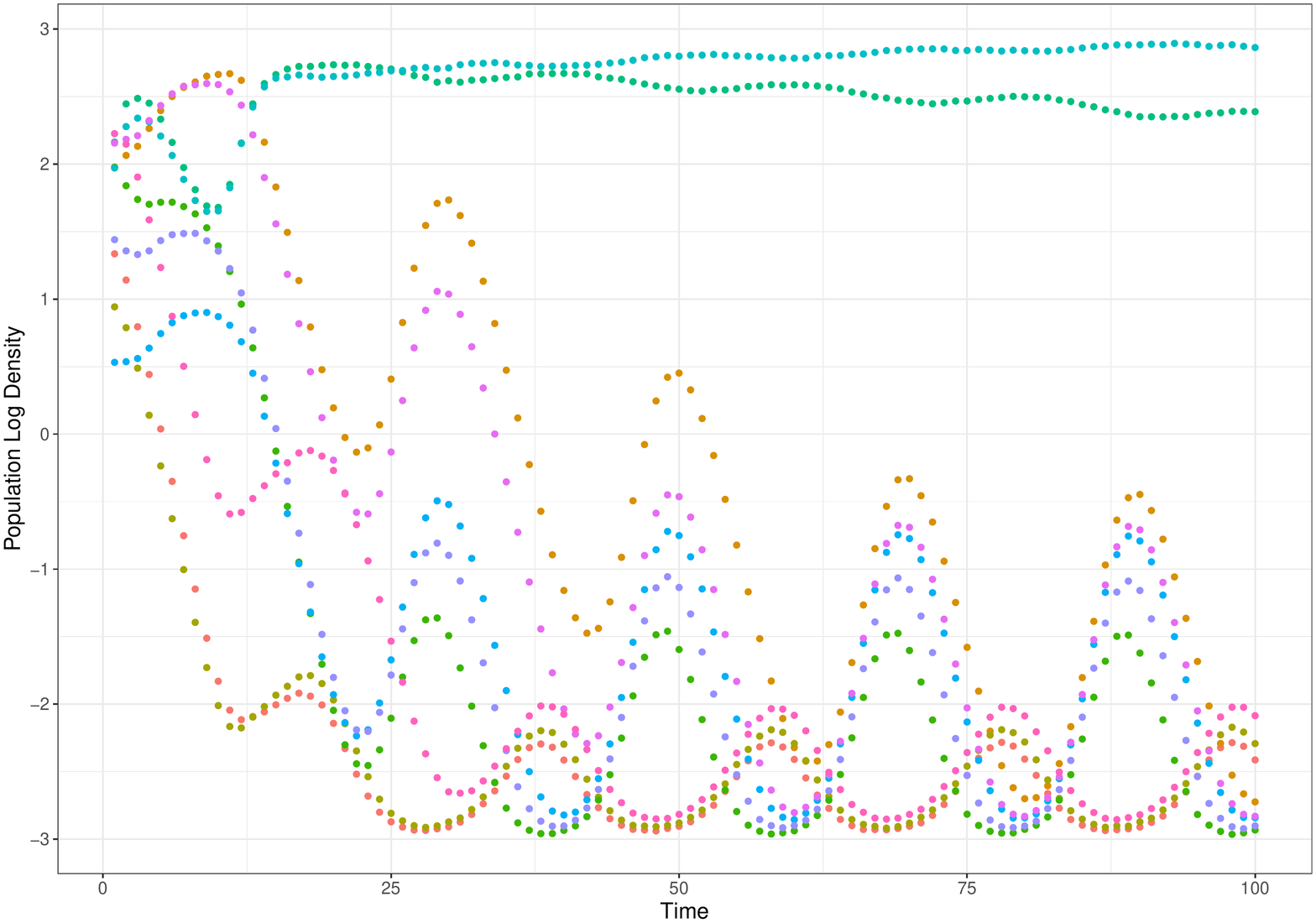}
\caption{Sample time series of the synthetic data. Each color represents a different population. In this simulation we choose $M=10,\sigma=0.01$}
\end{figure}

\begin{figure}[H]
\centering
\includegraphics[width=\textwidth]{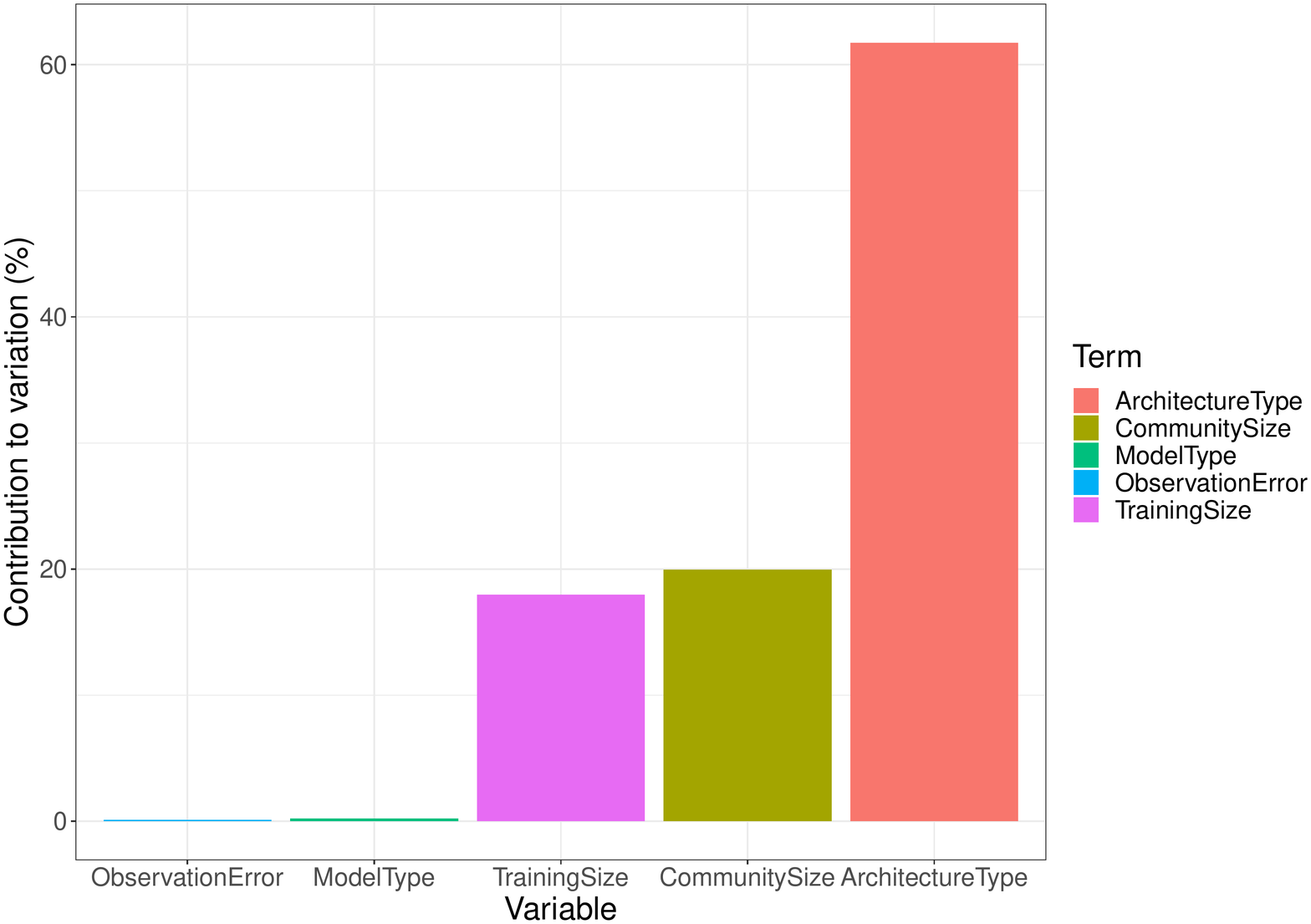}
\caption{Contribution to the variation of the RMSE of the different variables included in our data analysis analyzed through an ANOVA.}
\end{figure}

\begin{figure}[H]
\centering
\includegraphics[width=\textwidth]{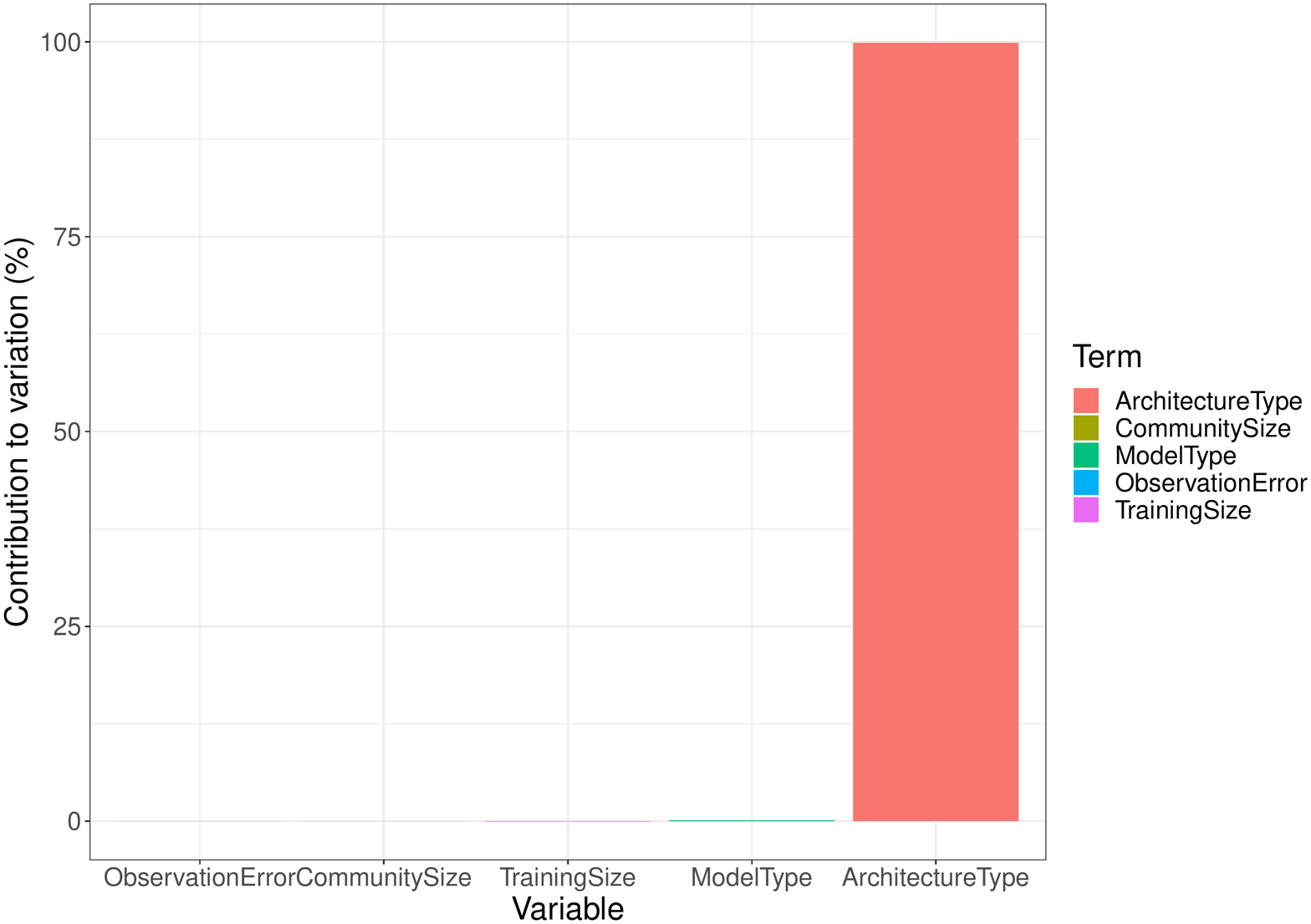}
\caption{Contribution to the variation of the size of the prediction intervals of the different variables included in our data analysis analyzed through an ANOVA.}
\end{figure}

\begin{figure}[H]
\centering
\includegraphics[width=\textwidth]{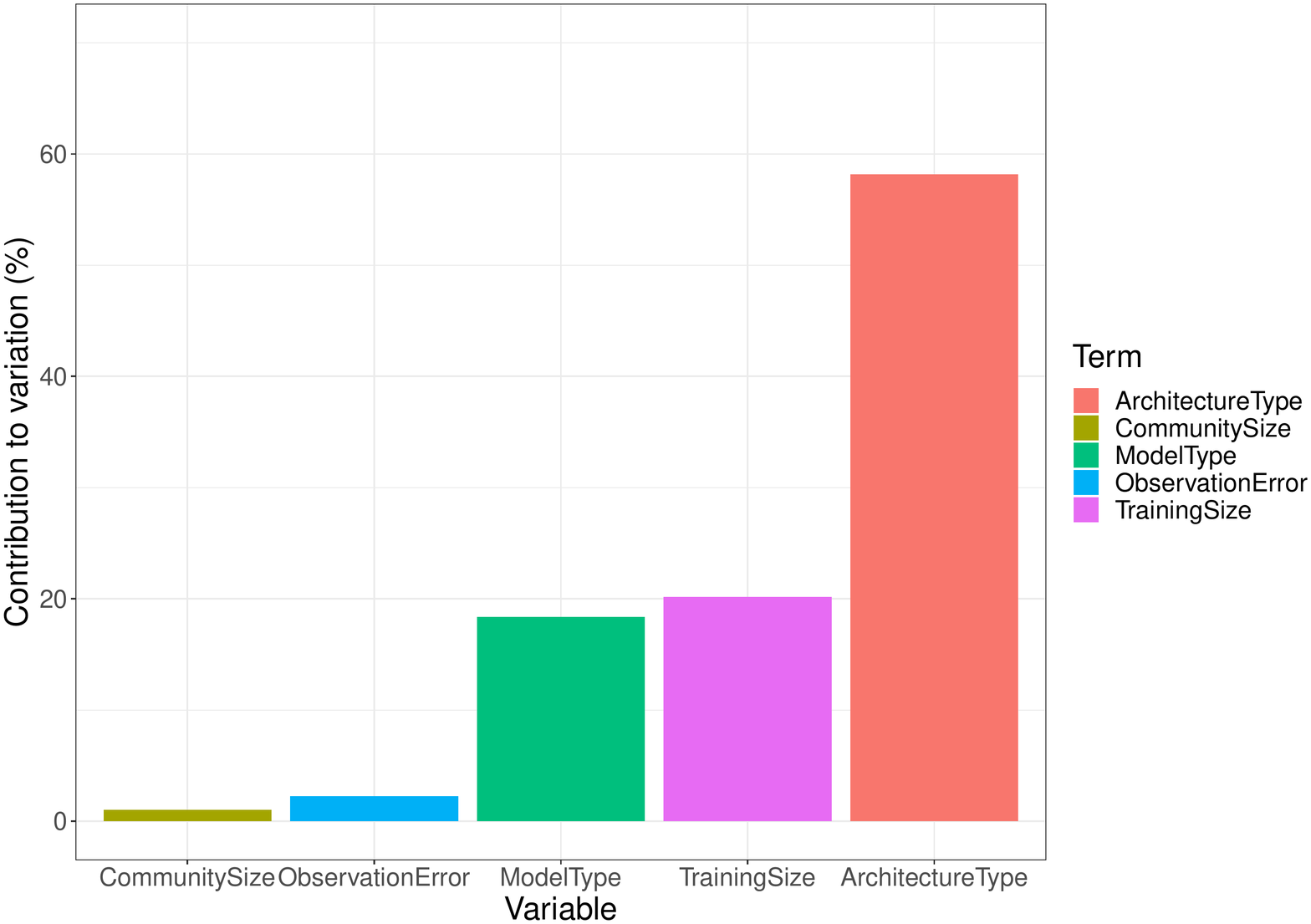}
\caption{Contribution to the variation of the interval scores of the different variables included in our data analysis analyzed through an ANOVA.}
\end{figure}

\begin{figure}[H]
\centering
\includegraphics[width=\textwidth]{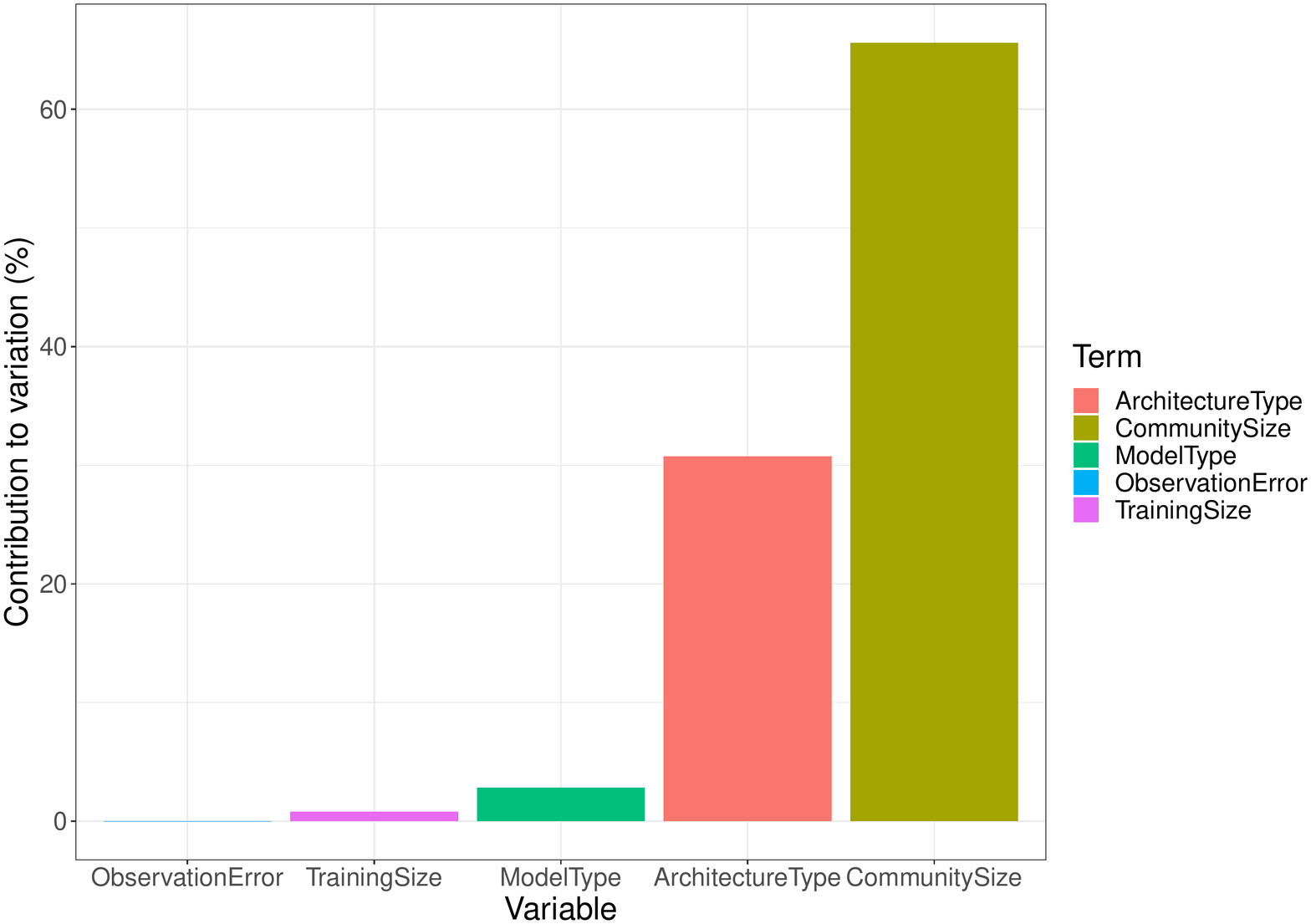}
\caption{Contribution to the variation of the ratio between the predicted final community size versus real final community size of the different variables included in our data analysis analyzed through an ANOVA.}
\end{figure}


\end{document}